\documentclass[preprint,aps,prb]{revtex4}
\usepackage{epsfig}
\usepackage{amsmath}
\usepackage{graphics}
\begin{document}

\title{Plane wave/pseudopotential implementation of 
excited state gradients in density functional 
linear response theory: a new route via implicit differentiation}
\author{ Nikos L. Doltsinis$^1$
and D. S. Kosov$^2$}
\address{$^1$Lehrstuhl f\"ur Theoretische Chemie, Ruhr-Universit\"at
Bochum, D-44780 Bochum, Germany\\
$^2$Department of Chemistry and Biochemistry, University of Maryland,
College Park, MD 20742 \\
}
\begin{abstract}
This work presents the formalism and implementation of
excited state nuclear forces
within density functional linear response theory (TDDFT)
using a plane wave basis set. 
An implicit differentiation technique is developed for computing 
nonadiabatic coupling between Kohn-Sham molecular orbital 
wavefunctions as well as gradients of orbital energies  
which are then used to calculate excited state nuclear forces.
The algorithm has been implemented in a plane wave/pseudopotential code
taking into account only a reduced active subspace of
molecular orbitals.  
It is demonstrated for the H$_2$ and N$_2$ molecules that
the analytical gradients rapidly converge to the exact
forces when the active subspace  
of molecular orbitals approaches completeness.
\\
\\ 
\\
{\bf \it J. Chem. Phys. 122, 144101 (2005)}
\end{abstract}
\maketitle

\section{Introduction}
The past decade has seen time-dependent density functional linear response
theory (TDDFT) \cite{rg84,c95,ba96} become the most widely used electronic
structure method for
calculating vertical electronic excitation energies \cite{mbagl02,th00}.
Except for certain well-known problem cases such as, for instance, charge
transfer \cite{bsh04,dwh03,cgggsd00} and double excitations \cite{hh99}, TDDFT excitation energies are generally
remarkably accurate, typically to within a fraction of an
electron Volt \cite{agb03,sggb00,hgr00,th98}.

Excited state analytical nuclear forces within TDDFT
have only been implemented recently \cite{ca99,ca00,fa02,h03} in an attempt to
extend the applicability
of TDDFT beyond single point calculations. One complication
has been the fact that 
TDDFT merely provides excitation energies, but excited state
wavefunctions are not 
properly defined. The first excited state geometry optimization using
analytical gradients 
was presented by van Caillie and Amos based on a Handy-Schaefer Z-vector method
\cite{ca99,ca00}.  
An extended Lagrangian
ansatz was chosen by Furche and Ahlrichs \cite{fa02} and Hutter \cite{h03} for
their Gaussian-type basis set and plane wave/pseudopotential implementations,
respectively. The latter variant is of particular importance
for condensed phase 
applications since it is used in conjunction with periodic boundary conditions.
In order to ensure completeness, the number of Kohn-Sham (KS) orbitals included
in constructing the response matrix in a molecular orbital
(MO) basis must equal the 
number of basis functions. Since a plane wave basis typically consists of two
orders of magnitude more basis functions than a
Gaussian-type basis set a complete MO 
formulation of TDDFT is impractical. A solution to this
problem is to cast the working  
matix equations directly into a plane wave basis as proposed
by Hutter \cite{h03}. Earlier, 
Doltsinis and Sprik \cite{ds00} have proposed an
alternative, {\it active space} approach  
to TDDFT in which only a subset of (active) KS orbitals is
selected to construct the 
response matrix. For a large variety of excited states,
convergence of the corresponding 
excitation energies has been shown to be rapid with respect
to the number of orbitals 
included in the active space \cite{ds00,cgggsd00}. In the
present paper, we shall follow this 
active space ansatz and derive analytical expressions for
excited state nuclear forces within 
an MO basis. In contrast to previous work, we do not rely on
a Lagrangian formulation \cite{fa02,h03,bk99}, 
but employ an implicit differentiation scheme instead. This
has the advantage that we obtain, 
in addition to the excited state energy gradients, also
the gradients of KS energies and 
wavefunctions. The latter may be exploited to compute
nonadiabatic coupling matrix elements 
between different electronic states.
We have implemented the working equations within a plane wave/pseudopotential
code \cite{cpmd} and we will demonstrate numerical accuracy 
and illustrate the convergence behaviour  with respect to
the number of MOs included in the active space 
 for some prototypical test examples. 

\section{Theory}
The linear response TDDFT eigenvalue problem can be written in
Hermitian form as
\begin{equation} \label{eigen}
{\bf \Omega} |{\bf F}_n\rangle=\omega_n^2|{\bf F}_n\rangle\quad,
\end{equation} 
where the response matrix ${\bf \Omega}$ is defined as
\begin{equation}\label{omega}
\Omega_{ph\sigma,p'h'\sigma'}=\delta_{\sigma\sigma'}\delta_{pp'}
\delta_{hh'}(\epsilon_{p\sigma}-\epsilon_{h\sigma})^2 + 
2\sqrt{\epsilon_{p\sigma}-\epsilon_{h\sigma}}K_{ph\sigma,p'h'\sigma'}
\sqrt{\epsilon_{p'\sigma}-\epsilon_{h'\sigma}}\quad,
\end{equation}
with the coupling matrix 
\begin{equation}
K_{ph\sigma,p'h'\sigma'}=\int d{\bf r}\int d{\bf r^{\prime}}
\psi_{p\sigma}({\bf r})\psi_{h\sigma}({\bf r})f_{\rm H,xc}^{\sigma\sigma'}({\bf r,
r^{\prime}}) \psi_{p'\sigma'}({\bf r^\prime})\psi_{h'\sigma'}({\bf
r^\prime})\quad,
\end{equation}
$\psi_{p\sigma}$ and $\psi_{h\sigma}$ being the
KS particle (unoccupied) and hole (occupied) molecular orbitals
with spin $\sigma$
corresponding to the KS energies $\epsilon_{p\sigma}$ and
$\epsilon_{h\sigma}$, respectively. The response kernel
\begin{equation}
f_{\rm H,xc}^{\sigma\sigma'}({\bf r,
r^{\prime}})=\frac{1}{|{\bf r -r^{\prime}}|}+\delta({\bf r}-{\bf
r}')\frac{\delta ^2E_{\rm xc}}{\delta\rho^\sigma({\bf r})\delta\rho^{\sigma^\prime}({\bf r}^\prime)}
\end{equation}
containing a Hartree term and an exchange-correlation term is given in
the usual adiabatic approximation \cite{gk90}, i.e. the
exchange-correlation contribution is taken to be simply the second
derivative of the static ground state exchange-correlation energy,
$E_{\rm xc}$, with respect to the spin density $\rho^\sigma$.

For the sake of simplicity, the following derivation is for singlet
excitations only (extension to triplet excitations is
straightforward). We shall therefore drop the spin index $\sigma$. The
reduced (singlet) response matrix is given by
\begin{equation}
\Omega_{ph,p'h'}=\delta_{pp'}
\delta_{hh'}(\epsilon_p-\epsilon_h)^2 + 
4\sqrt{\epsilon_p-\epsilon_h}K_{ph,p'h'}
\sqrt{\epsilon_{p'}-\epsilon_{h'}}\quad,
\end{equation}
where
\begin{equation}
K_{ph,p'h'}=\int d{\bf r}\int d{\bf r^{\prime}}
\psi_p({\bf r})\psi_h({\bf r})f_{\rm H,xc}({\bf r,
r^{\prime}}) \psi_{p'}({\bf r^\prime})\psi_{h'}({\bf
r^\prime})\quad,
\end{equation}
and
\begin{equation}
f_{\rm H,xc}({\bf r,
r^{\prime}})=\frac{1}{|{\bf r -r^{\prime}}|}+\delta({\bf r}-{\bf
r}')\frac{\delta ^2E_{\rm xc}}{\delta\rho({\bf r})^2}\quad.
\end{equation}

Multiplying eq.~(\ref{eigen}) by $\langle {\bf F}_n|$ from the left we
obtain
\begin{equation} \label{eigen2}
\langle {\bf F}_n|{\bf \Omega} |{\bf F}_n\rangle=\omega_n^2\quad.
\end{equation} 
Differentiation with respect to the nuclear coordinate
$R_\alpha \quad (\alpha=1,\dots,3N)$ for a molecule consisting
of $N$ atoms yields 
\begin{equation} \label{exc-grad}
\omega_n^\alpha=\frac{1}{2\omega_n}\langle {\bf F}_n|{\bf
\Omega}^\alpha |{\bf
F}_n\rangle=\frac{1}{2\omega_n}\sum_{ph}\sum_{p'h'}F_{ph}^{(n)}\Omega_{ph,p'h'}^\alpha
F_{p'h'}^{(n)}\quad,
\end{equation} 
where we have used the short-hand notation $\frac{d\, f}{dR_\alpha}\equiv f^\alpha$; the $F_{ph}^{(n)}$ are the components of the linear response
eigenvector ${\bf F}_n$. Carrying out the differentiation of the
response matrix, eq.~(\ref{exc-grad}) becomes
\begin{eqnarray} \label{exc-grad2}\nonumber
\omega_n^\alpha & = & 
\frac{1}{\omega_n}\left[\sum_{ph}(F_{ph}^{(n)})^2
(\epsilon_p^\alpha-\epsilon_h^\alpha)(\epsilon_p-\epsilon_h)\right.\\\nonumber
&&+ 2\int d{\bf r}\int  d{\bf r}' \Gamma_1({\bf r})f_{\rm
H,xc}({\bf r,r^{\prime}})\Gamma_2({\bf r}')\\
&&\left.+ 2\int d{\bf r} \Gamma_1({\bf r})
\frac{\delta ^3E_{\rm xc}}{\delta\rho({\bf r})^3}
\rho^\alpha({\bf r})\Gamma_1({\bf r})\right]
\quad.
\end{eqnarray} 
Here we have defined the contracted densities
\begin{equation} \label{gamma1}
\Gamma_1({\bf r})=
\sum_{ph}F_{ph}^{(n)}\sqrt{\epsilon_p-\epsilon_h}\Gamma_{ph}({\bf r}) 
\end{equation} 
and
\begin{equation} \label{gamma2}
\Gamma_2({\bf r})=
\sum_{ph}F_{ph}^{(n)}\left[
\frac{\epsilon_p^\alpha-\epsilon_h^\alpha}{\sqrt{\epsilon_p-\epsilon_h}}
\Gamma_{ph}({\bf r}) 
+2\sqrt{\epsilon_p-\epsilon_h}\Gamma_{ph}^\alpha({\bf r})\right]
\end{equation} 
with
\begin{equation} \label{gamma-ph}
\Gamma_{ij}({\bf r})=\psi_i({\bf r})\psi_j({\bf r})
\end{equation} 
In order to compute the excitation energy gradient
(eq.~(\ref{exc-grad2})), we require the nuclear derivatives of KS
orbital energies and 
wavefunctions, $\epsilon_i^\alpha$ and $\psi_i^\alpha$
($i=p,h$). These can be obtained using an implicit differentiation
scheme as follows. We start by writing down the KS equations
in matrix form
\begin{equation} \label{ks-eq}
F_{ij}\equiv H_{ij}-\epsilon_i\delta_{ij}=0\quad.
\end{equation} 
For the full differential of $F_{ij}$ we have
\begin{equation} \label{ks-eq-diff}
dF_{ij}=(\frac{\partial H_{ij}}{\partial R_{\alpha}}-\epsilon_i^\alpha\delta_{ij})dR_\alpha + \sum_k \int d{\bf
r}H_{ij}^k\delta\psi_k({\bf r}) =0\quad,
\end{equation} 
where $H_{ij}^k \equiv \frac{\delta H_{ij}}{\delta\psi_k({\bf
r})}$. Division by $dR_\alpha$ yields
\begin{equation} \label{ks-eq-diff2}
\frac{\partial H_{ij}}{\partial R_{\alpha}}-\epsilon_i^\alpha\delta_{ij}
=- \sum_k \int d{\bf r}H_{ij}^k\psi_k^\alpha({\bf r})
= - \sum_k \int d{\bf r}\int d{\bf r}'H_{ij}^k \delta({\bf r}-{\bf r}')\psi_k^\alpha({\bf r}')\quad.
\end{equation} 
On the rhs of eq.~(\ref{ks-eq-diff2}) we have inserted a delta
function, which we now express in terms of KS orbitals
\begin{equation} \label{delta}
\delta({\bf r}-{\bf r}')=\sum_l\psi_l({\bf r})\psi_l({\bf r}')
\quad.
\end{equation} 
Thus eq.~(\ref{ks-eq-diff2}) becomes
\begin{equation} \label{ks-eq-diff3}
\frac{\partial H_{ij}}{\partial R_{\alpha}}-\epsilon_i^\alpha\delta_{ij}
=- \sum_{kl} H_{ij}^{kl}\psi_k^{\alpha l}
\quad,
\end{equation} 
where 
\begin{eqnarray} \label{hklij}
\nonumber
H_{ij}^{kl}&\equiv& \int d{\bf r}H_{ij}^k\psi_l({\bf r})\\
&=&(\delta_{ik} \delta_{lj} + \delta_{jk} \delta_{li}) 
\epsilon_l + 
 2 n_k \int d{\bf r}\int d{\bf r'} \Gamma_{kl} ({\bf r}) 
f_{\rm H, xc}({\bf r}, {\bf r}') \Gamma_{ij} ({\bf r'}) 
\quad,
\end{eqnarray} 
and 
\begin{equation} \label{nonadiab}
\psi_k^{\alpha l}\equiv \int d{\bf r}\psi_l({\bf r})\psi_k^\alpha({\bf r})
\quad,
\end{equation} 
$n_k$ being the number of electrons occupying orbital $k$.

Exploiting the symmetry of the nonadiabatic coupling matrix elements
(\ref{nonadiab}), i.e. $\psi_l^{\alpha k}=-\psi_k^{\alpha l}$ and
therefore $\psi_l^{\alpha l}=0$,
eq.~(\ref{ks-eq-diff3}) can be rewritten as
\begin{equation} \label{ks-eq-diff4}
\frac{\partial H_{ij}}{\partial R_{\alpha}}=\sum_{l<k}D_{ij}^{lk}\psi_k^{\alpha l}
\quad,(i<j)
\end{equation} 
and for the diagonal terms ($i=j$)
\begin{equation} \label{ks-energy-grad}
\epsilon_i^\alpha=\frac{\partial H_{ii}}{\partial R_{\alpha}}-\sum_{k<l}D_{ii}^{kl}
\psi_k^{\alpha l} 
\quad.
\end{equation} 
where
\begin{equation} \label{dmat}
D_{ij}^{lk}=H_{ij}^{lk}-H_{ij}^{kl}=
(\delta_{il}\delta_{kj}+\delta_{ik}\delta_{lj})(\epsilon_k-\epsilon_l)
+2 (n_k-n_k) K_{ij,lk}
\end{equation} 
With the definition (\ref{dmat}) eq. (\ref{ks-eq-diff4}) becomes
\begin{equation} \label{ks-eq-diff5}
\frac{\partial H_{hp}}{\partial R_{\alpha}}=
\sum_{p'h'}((\epsilon_{p'}-\epsilon_{h'})\delta_{pp'} \delta_{hh'}+4 K_{p'h',ph}) \psi_{h'}^{\alpha p'}
\end{equation} 
for particle-hole states, and
\begin{equation} \label{ks-eq-diff6}
\frac{\partial H_{ij}}{\partial R_{\alpha}}=
4 \sum_{ph}K_{ij,ph} \psi_h^{\alpha p}
+ (\epsilon_i-\epsilon_j)\psi_i^{\alpha j}
\quad,(i<j,\; ij\ni ph)
\end{equation} 
for all remaining combinations. Eq. (\ref{ks-eq-diff6}) allows us to
express the nonadiabatic coupling elements between non-particle-hole
states analytically as
\begin{equation} \label{non-ph-coupl}
\psi_i^{\alpha j}=\frac{
\frac{\partial H_{ij}}{\partial R_{\alpha}}+
4 \sum_{ph} K_{ij, ph}\psi_h^{\alpha p}}{
(\epsilon_i-\epsilon_j)}
\quad,(i<j,\; ij\ni ph)
\end{equation} 
The system of linear equations (\ref{ks-eq-diff5}) is first solved for
the particle-hole nonadiabatic coupling elements $\psi_p^{\alpha h}$,
which are then  
inserted into eq.~(\ref{non-ph-coupl}) to obtain the remaining,
non-particle-hole, elements. The second term in the
numerator of eq.~(\ref{non-ph-coupl}) is most conveniently
evaluated by introducing the contracted density 
\begin{equation} \label{gamma3}
\Gamma_3({\bf r})=
\sum_{ph} \psi_p({\bf r})\psi_h({\bf r})\psi_h^{\alpha p}
\end{equation} 
Then
\begin{equation} \label{sum_ph}
\sum_{ph} K_{ij, ph}\psi_p^{\alpha h}=
\int d{\bf r}\int d{\bf r^{\prime}}
\Gamma_3({\bf r})f_{\rm H,xc}({\bf r,
r^{\prime}}) \psi_i({\bf r^\prime})\psi_j({\bf
r^\prime})\equiv K_{ij}'
\end{equation} 
Thus eq.~(\ref{non-ph-coupl}) becomes
\begin{equation} \label{non-ph-coupl2}
\psi_j^{\alpha i}=\frac{
\frac{\partial H_{ij}}{\partial R_{\alpha}}+
4 K_{ij}'}
{(\epsilon_i-\epsilon_j)}
\quad,(i<j,\; ij\ni ph)
\end{equation} 
Similarly the KS orbital
energy gradients can now be obtained from the simplified
eq.~(\ref{ks-energy-grad}) 
\begin{equation} \label{ks-energy-grad2}
\epsilon_i^\alpha=\frac{\partial H_{ii}}{\partial R_{\alpha}}+4K_{ii}'
\quad.
\end{equation} 
Finally, the nuclear derivative of the KS orbital
wavefunction is recovered by unfolding the nonadiabatic couplings
\begin{equation} \label{orb-grad}
\psi_k^\alpha({\bf r})=\sum_l\psi_l({\bf r})\psi_k^{\alpha l}
\quad.
\end{equation} 

Equations (1)--(31) have been implemented with periodic boundary conditions
using a plane wave expansion of the KS MOs at the $\Gamma$
point of the Brillouin zone. By making use of
the periodic boundary conditions, the generalized densities 
$\Gamma_1$, $\Gamma_2$, $\Gamma_3$ and $\Gamma_{ij}$
can be expanded
in reciprocal space via the three-dimensional Fourier transform, e.g.
\begin{equation}
\Gamma_1({\bf r}) =  \sum_{ {\bf G}} \Gamma_1({\bf G}) \exp(i{\bf Gr})
\end{equation}
where ${\bf G} $ is the vector of
the recipocal lattice. The Hartree part of the matrix element
$\int d{\bf r}\int  d{\bf r}' \Gamma_1({\bf r})f_{\rm
H,xc}({\bf r,r^{\prime}})\Gamma_2({\bf r}')$ 
and $\int d{\bf r}\int d{\bf r'} \Gamma_{kl} ({\bf r}) 
f_{\rm H, xc}({\bf r}, {\bf r}') \Gamma_{ij} ({\bf r'})$
which enter the key equations (\ref{exc-grad2}) 
and (\ref{hklij}), respectively, can be readily computed
in reciprocal space, e.g.
\begin{equation} 
\int d{\bf r}\int  d{\bf r}' \Gamma_1({\bf r})\frac{1}{|{\bf
r}-{\bf r'}|}\Gamma_2({\bf r}')=
\Omega \sum_{{\bf G}\ne 0} \frac{2 \pi}{G^2} \Gamma_1({\bf G}) \Gamma_2({\bf G}) 
\end{equation}
whereas the exchange-correlation parts of the matrix elements are
calculated via direct numerical integration over grid in coordinate space.

\section{Test results}
To illustrate the performance and the convergence behaviour
of our method, we have computed nuclear gradients of the first excited state 
energies of H$_2$ and N$_2$.  
The calculations were performed using our implementation of
the formalism presented here in the 
 CPMD package \cite{cpmd}. All the systems
were treated employing periodic boundary conditions and the molecular orbitals 
were expanded in plane waves at the $\Gamma$ point of the Brillouin zone.
We used Troullier-Martins normconserving pseudopotentials \cite{tm91}.
The excitation energies and nuclear gradients were computed  within the 
adiabatic local density approximation \cite{pgg96} to the linear
response exchange-correlation kernel.

The central idea underlying the present active space
approach originates from the observation that excitation
energies for a large number of electronic transitions
exhibit only a minor dependence on the size of the response
matrix (\ref{omega}). This is illustrated in Table~1 for the
$3\sigma_g \rightarrow 1\pi_g$ transition of
N$_2$. A simple two-state HOMO--LUMO response calculation is
seen to give an excitation energy which is less than 0.2~eV
away from an extended treatment including all 5 occupied and
100 virtual MOs.
Generally, such behaviour is to be expected for
excitations which can be characterized by only a few
low-lying one-electron transitions without higher-lying
continuum states mixing in.

In the following, we shall discuss the active space
dependence of the excitation energy gradients
 and the KS orbital energy gradients. The upper panel of Fig.~1
displays the completeness of the active space as a function of 
the number of virtual KS orbitals included in the space.
The integral  
\begin{equation}\label{complete}
C(N)= \int d{\bf r} \sum_{i=1}^N \psi_i({\bf r}) \psi_i(0)
\end{equation}
was used as a measure of completeness of the active space.
It becomes unity when the active space of KS orbitals is complete, i.e.
when the total number of the KS orbitals (virtual and occupied)
equals the number of plane waves used to solve the KS equations. 
The total number of plane waves is 925
for the 6 a.u. cubic box and 40 Ry plane wave cutoff. With 450 virtual orbitals
included the active space is almost complete and the
value of the integral (\ref{complete}) deviates from unity
by approximately $10^{-3}$ which is already comparable with the
accuracy of the numerical integration. 
The lower panel of Fig.~1 shows the
absolute deviation of the analytic derivatives from the
respective finite difference values for the
the first singlet excitation energy, $\omega_1$, as well as
the HOMO and LUMO KS orbital energies, $\epsilon_1$ and
$\epsilon_2$, of  H$_2$ at a bond length of 
1.0 a.u. as a function of the number of virtual KS orbitals
included in the active space. The absolute deviation in analytical gradients
vanishes rapidly as the number of virtual orbitals is increased 
and the errors in the analytical gradients of different
states generally show the same patterns
in the dependence upon the number of virtual orbitals included in the
active space.

Fig.~2 shows the
absolute deviation of the analytical derivative from the
respective finite difference value as a function of the size
of the active space for the first three KS orbital energies
$\epsilon_i$ ($i=1,2,3$)
as well as the lowest response matrix eigenvalue $\omega_1$
of the N$_2$ molecule. The errors of the analytical
gradients are seen to decrease rapidly as the number of
orbitals included in the active space
approaches the number of plane wave basis functions (in this
case 925 plane waves). For the largest active space the
deviations of all energies are of the order of $10^{-3}$ or
smaller. At this point, the accuracy of the analytical
derivatives is hard to assess because the finite difference
reference values are also subject to numerical errors. 
We have further checked whether the excited state
gradients are invariant under translation. The translational
contribution to the excitation energy gradient is found to
decrease rapidly as the active space increses. Interstingly,
the underlying ground state calculation exhibits a
significantly larger translational error than the excitation
energy.

To test the practical value of our derivatives, we have
performed geometry optimizations of $N_2$ in the first
excited state ($8\times 5.6 \times 5.6$~a.u. box, 40~Ry plane-wave
cutoff, i.e. 600 basis functions). When we include only 100 virtual orbitals in
the active space, we obtain a bond length of 2.44~a.u.,
which deviates by 0.02~a.u. from the value of
2.42~a.u. determined by a series of 
single point energy calculations. Upon increasing the number
of virtual states to 200, the optimized bond length comes
out as 2.42~a.u., correct to two decimal places. Our test
calculations illustrate how the size of the active space may
be adjusted to achieve any desired level
of accuracy. For many practical purposes it will be
sufficient to work with a reduced active space which is
significantly smaller than the total number of basis
functions. 

The situation is different in the case of molecular dynamics
simulations, where the nuclear forces need to be essentially
exact derivatives of the potential in order to
maintain conservation of energy. Although we have already
carried out test excited state MD simulations for
diatomic molecules, those results do not offer any
additional insight into the general performance and
convergence pattern of our method. 
We have not yet applied the formalism presented here to
perform more
realistic excited state MD simulations of polyatomic molecules
, because our current implementation does not
yet make use of more efficient iterative techniques, such as
Lanczos algorithm or related schemes \cite{ssf98},
to solve the response eigenvalue problem (\ref{eigen}). These
numerical techniques, however, are standard and we plan to
exploit them in future implementations. The scope of this
article is merely the presentation of the formalism and the
analysis of the convergence behaviour with respect to the
choice of the active space.

We would like to emphasize, however, that the method
described here is capable of providing additional
information beyond excited state energy gradients. Fig.~3
shows, for instance, the nonadiabatic coupling strength
between the second and third KS orbitals, $\psi_2$ and
$\psi_3$, for the H$_2$ molecule as a function of its bond
length. The nonadiabatic coupling values obtained from
eqn~(\ref{non-ph-coupl2}) exhibit a singularity at the
crossing point between the two KS orbital energies, as one
would expect due to the KS energy difference in the
denominator. This feature of our formalism may be exploited
in future applications of TDDFT beyond the Born-Oppenheimer
approximation.

\section{Conclusions}
We have developed and implemented an novel, alternative
formalism to calculate analytical
nuclear forces for TDDFT excited states within a plane 
wave/pseudopotential framework. In addition to the excited
state energy gradient, our method also provides molecular
orbital wavefunction as well as energy derivatives at a
small computational overhead compared to the vertical
excitation energies. The latter may, for instance be employed
as a powerful tool for understanding and interpreting
various chemical phenomena such as molecular structures and
reactivities \cite{yrgsf94}. 
A fundamental quantity in the present formalism are the
nonadiabatic coupling elements in the molecular orbital
basis which are obtained as direct solutions of a system of
linear equations. These matrix elements may provide the
basis for the calculation of nonadiabatic couplings
between the many-electron adiabatic wavefunctions. 
Trial calculations on the prototypical test molecules H$_2$
and N$_2$
demonstrate that our implementation reproduces the exact gradients 
when the number of molecular orbital included in the active
space approaches the number of plane wave basis functions. 
Excited state geometry optimization of N$_2$ using different
active spaces show that for many practical purposes it will
be sufficient to work within a relatively small active
space. The size of the latter may be tuned to achieve any
desired level of accuracy.

{\bf Acknowledgments} \\[-0.5cm]

We are grateful to J. Hutter and F. Furche for helpful
discussions.

\clearpage
\begin{appendix}
\section{Calculation of $\frac{\partial H_{ij}}{\partial R}$ in plane waves}

The  matrix elements of the derivatives of
the KS Hamiltonian  are required for the calculations of
the KS molecular orbitals gradients.
Since the kinetic and exchange-correlation energies do not depend 
directly upon the atomic positions, the matrix element of the
derivative becomes (local/nonlocal pseudopotential 
 and electrostatic interaction contributions):
\begin{equation}
\frac{\partial H_{ij}}{\partial R_I} =
\frac{\partial }{\partial R} H^{pp,local}_{ij} +
\frac{\partial }{\partial R} H^{pp,nonlocal}_{ij} +
\frac{\partial }{\partial R} H^{es}_{ij} 
\end{equation}
All these matrix elements are computed in reciprocal space.
The matrix element of the derivative of
the local pseudopotential has the following form
\begin{equation}
\frac{\partial }{\partial {\bf R}_I} H^{pp,local}_{ij}
=- \Omega \sum_{{\bf G}} i {\bf G} V_{local}^I ({\bf G}) S_I({\bf G}) \Gamma_{ij}({\bf G})
\end{equation}
where ${\bf R}_I$ denotes the atomic position and the structure factor $S_I= \exp(i \bf {G R}_I)$
of nuclei $I$, $\Gamma_{ij}({\bf G})$ is the
three-dimensional Fourier transform of the contracted density
(\ref{gamma-ph}). 
The nonlocal pseudopotential contribution
is
\begin{equation}
\frac{\partial }{\partial {\bf R}_I} H^{pp,nonlocal}_{ij}=
\sum_{\mu \nu \in I} \left[ (
\frac{ \partial F^{\mu}_{I,i} }{ \partial {\bf R}_I})^* h^I_{\mu \nu} F^{\nu}_{I,j}
+ (F^{\mu}_{I,i})^* h^I_{\mu \nu} \frac{\partial F^{\nu}_{I,j}}{\partial {\bf R}_I}
\right] \;,
\end{equation}
where the contribution from the projector derivative 
$\frac{\partial F^{\mu}_{I,i} }{ \partial {\bf R}_I}$
is calculated in the standard way \cite{mh00}.
The contribution from the electrostatic energy is computed in the form
\begin{equation}
\frac{\partial }{\partial {\bf R}_I} H^{es}_{ij}=
- \Omega \sum_{{\bf G} \ne 0} i \frac{{\bf G}}{G^2} \Gamma^*_{ij} ( {\bf G}) n^I_c ({\bf G})
 S_I({\bf G})
\end{equation}
where  $n^I_c ({\bf G})$ is  the gaussian core charge distribution for 
nuclei $I$ in reciprocal space.

\section{Third functional derivatives of the exchange-correlation functional}
\subsection{Third derivative of Vosko-Wilk-Nusair correlation}%
\begin{equation}
E_c=\int \rho \epsilon_c d{\bf r}\quad,
\end{equation}
where
\begin{equation}
\epsilon_c= A \left\{ \ln{\frac{x^2}{X(x)}} +
\frac{2b}{Q}
\tan^{-1}{\frac{Q}{X'(x)}}
-\frac{bx_0}{X(x_0)}
\left[ \ln{\frac{(x-x_0)^2}{X(x)}}+
\frac{2X'(x_0)}{Q}\tan^{-1}{\frac{Q}{X'(x)}}
\right]\right\} 
\end{equation}
with $x=\sqrt{r_s}$, $r_s=(\frac{3}{4\pi \rho})^{\frac{1}{3}}$,
$X(x)=x^2+bx+c$, $X'(x)\equiv\frac{dX}{dx}=2x+b$, 
$Q=\sqrt{4c-b^2}$, $A=0.0310907$, $b=3.72744$, $c=12.9352$, $x_0=-0.10498$.
\begin{equation}\label{dEdrho}
\frac{\delta E_c}{\delta \rho}=\epsilon_c+\rho\frac{\partial
\epsilon_c}{\partial \rho}
\end{equation}
where
\begin{equation}\label{dedrho}
\frac{\partial \epsilon_c}{\partial \rho}=\frac{\partial
x}{\partial \rho}\frac{\partial \epsilon_c}{\partial x}= 
-\frac{x}{6\rho}\frac{\partial \epsilon_c}{\partial x}
\end{equation}
and 
\begin{equation}
\frac{\partial \epsilon_c}{\partial
x}=A\left\{\frac{2}{x}-\frac{X}{X'}-\frac{4b}{X'^2+Q^2}-\frac{bx_0}{X(x_0)}\left[\frac{2}{x-x_0}-\frac{X}{X'}-\frac{4X'(x_0)}{X'^2+Q^2}\right]\right\}
\end{equation}
\begin{equation}\label{d2Edrho2}
\frac{\delta^2 E_c}{\delta \rho^2}=2\frac{\partial
\epsilon_c}{\partial \rho}+\rho\frac{\partial^2
\epsilon_c}{\partial \rho^2}
\end{equation}
with
\begin{eqnarray}\label{d2edrho2}
\frac{\partial^2 \epsilon_c}{\partial \rho^2}&=&\frac{\partial}{\partial
\rho}\left(-\frac{1}{6}\frac{x}{\rho}\frac{\partial
\epsilon_c}{\partial x}\right) 
=-\frac{1}{6}\left[ \left(\frac{\partial}{\partial
\rho}\frac{x}{\rho}\right)\frac{\partial \epsilon_c}{\partial x} 
+ \frac{x}{\rho}\left(\frac{\partial}{\partial \rho}\frac{\partial
\epsilon_c}{\partial x}\right) \right]\\
 &=&\frac{7}{36}\frac{x}{\rho^2}\frac{\partial
\epsilon_c}{\partial x}+\frac{1}{36}\frac{x^2}{\rho^2}\frac{\partial^2
\epsilon_c}{\partial x^2}
\end{eqnarray}
Using relation (\ref{dedrho}), eqn (\ref{d2edrho2}) can be rewritten
as
\begin{equation}\label{d2edrho2f}
\frac{\partial^2 \epsilon_c}{\partial \rho^2}=-\frac{7}{6}\frac{1}{\rho}\frac{\partial
\epsilon_c}{\partial \rho}+\frac{1}{36}\frac{x^2}{\rho^2}\frac{\partial^2
\epsilon_c}{\partial x^2}
\end{equation}
and thus eqn (\ref{d2Edrho2}) becomes
\begin{equation}
\frac{\delta^2 E_c}{\delta \rho^2}=\frac{5}{6}\frac{\partial
\epsilon_c}{\partial \rho}+\frac{1}{36}\frac{x^2}{\rho}\frac{\partial^2
\epsilon_c}{\partial x^2}
\end{equation}
where
\begin{eqnarray}
\frac{\partial^2 \epsilon_c}{\partial x^2}&=& A\left\{
-\frac{2}{x^2}-\frac{2}{X}+\left(\frac{X'}{X}\right)^2+\frac{16bX'}{(X'^2+Q^2)^2}\right.\\
&&\left.+\frac{bx_0}{X(x_0)}\left[\frac{2}{(x-x_0)^2}+\frac{2}{X}-\left(\frac{X'}{X}\right)^2-\frac{16X'(x_0)X'(x)}{(X'^2+Q^2)^2}\right]\right\}
\end{eqnarray}
\begin{eqnarray}\label{d3Edrho3}
\frac{\delta^3 E_c}{\delta \rho^3}&=&3\frac{\partial^2
\epsilon_c}{\partial \rho^2}+\rho\frac{\partial^3
\epsilon_c}{\partial \rho^3}\\
\label{d3Edrho3f}&=&-\frac{1}{36}\left[35 
\frac{\partial\epsilon_c}{\partial\rho}+ 
\frac{1}{2}\frac{x^2}{\rho}\frac{\partial^2
\epsilon_c}{\partial x^2}+\frac{1}{6}\frac{x^3}{\rho}\frac{\partial^3
\epsilon_c}{\partial x^3}\right]
\end{eqnarray}
with
\begin{eqnarray}
\frac{\partial^3 \epsilon_c}{\partial x^3}&=&
A\left\{\frac{4}{x^3}+6\frac{X'}{X^2}-2\left(\frac{X'}{X}\right)^3+
\frac{32b}{(X'^2+Q^2)^2} \left[1-\frac{4X'^2}{X'^2+Q^2}\right]\right.
\\
\nonumber &&\left.+\frac{bx_0}{X(x_0)}\left[-\frac{4}{(x-x_0)^3}-6\frac{X'}{X^2}+2
\left(\frac{X'}{X}\right)^3-\frac{32X'(x_0)}{(X'^2+Q^2)^2} \left[
1-\frac{4X'^2}{X'^2+Q^2}\right]\right] \right\}
\end{eqnarray}
One arrives at eqn (\ref{d3Edrho3f}) by substituting eqn
(\ref{d2edrho2f}) into eqn (\ref{d3Edrho3}). For instance,
\begin{eqnarray}
\frac{\partial^3\epsilon_c}{\partial\rho^3}&=&\frac{\partial}{\partial\rho}
\left[-\frac{7}{6}\frac{1}{\rho}\frac{\partial
\epsilon_c}{\partial \rho}+\frac{1}{36}\frac{x^2}{\rho^2}\frac{\partial^2
\epsilon_c}{\partial x^2}\right]\\
&=&-\frac{7}{6}\frac{1}{\rho}\left(\frac{\partial^2
\epsilon_c}{\partial \rho^2}-\frac{1}{\rho}\frac{\partial
\epsilon_c}{\partial
\rho}\right)-\frac{1}{108}\frac{x^2}{\rho^3}\left(
7\frac{\partial^2\epsilon_c}{\partial x^2}+\frac{1}{2}x\frac{\partial^3
\epsilon_c}{\partial x^3}\right)
\end{eqnarray} 
where we have used
\begin{equation}
\frac{\partial}{\partial\rho}\left(\frac{x}{\rho}\right)^2=-\frac{7}{3}\frac{x^2}{\rho^3}
\end{equation}
\subsection{Third derivative of Slater exchange}
\begin{equation}
E_x=\int \rho \epsilon_x d{\bf r}\quad,
\end{equation}
where
\begin{equation}
\epsilon_x=C\alpha\rho^{\frac{1}{3}}\; , \quad
C=-\frac{9}{8}\left(\frac{3}{\pi}\right)^{\frac{1}{3}}  \; , \quad \alpha=\frac{2}{3}
\end{equation}
The third derivatives can be readily computed
\begin{equation}\label{d3Exdrho3}
\frac{\delta^3 E_x}{\delta \rho^3}=-C\alpha^4\rho^{-\alpha-1}
\end{equation}
\noindent  
\end{appendix}

\clearpage
\bibliography{dfg}
\bibliographystyle{jcp}

\clearpage
\begin{table}[H]
\begin{center}
\begin{tabular}{|c|cccccc|}\hline
active space& $\Delta E_{\rm KS}$ & 1o/1v & 5o/1v&
5o/5v&5o/50v&5o/100v\\\hline\hline
exc. energy & 8.39 & 9.47 & 9.40 & 9.40 & 9.33 & 9.29\\\hline
\end{tabular}
\caption{Dependence of N$_2$ TDLDA excitation
energy ($^1\!\Pi_g$, $3\sigma_g \rightarrow 1\pi_g$) in eV  using
plane waves (p.w.) with a 70~Ry cutoff in a
10~$a_0$ periodic box on the number of occupied (o) and virtual (v)
Kohn-Sham orbitals included in 
the active space. $\Delta E_{\rm KS}$ is the unperturbed Kohn-Sham
energy difference.}
\end{center}
\end{table}
\clearpage

\clearpage

\begin{figure}
\centerline{
\epsfig{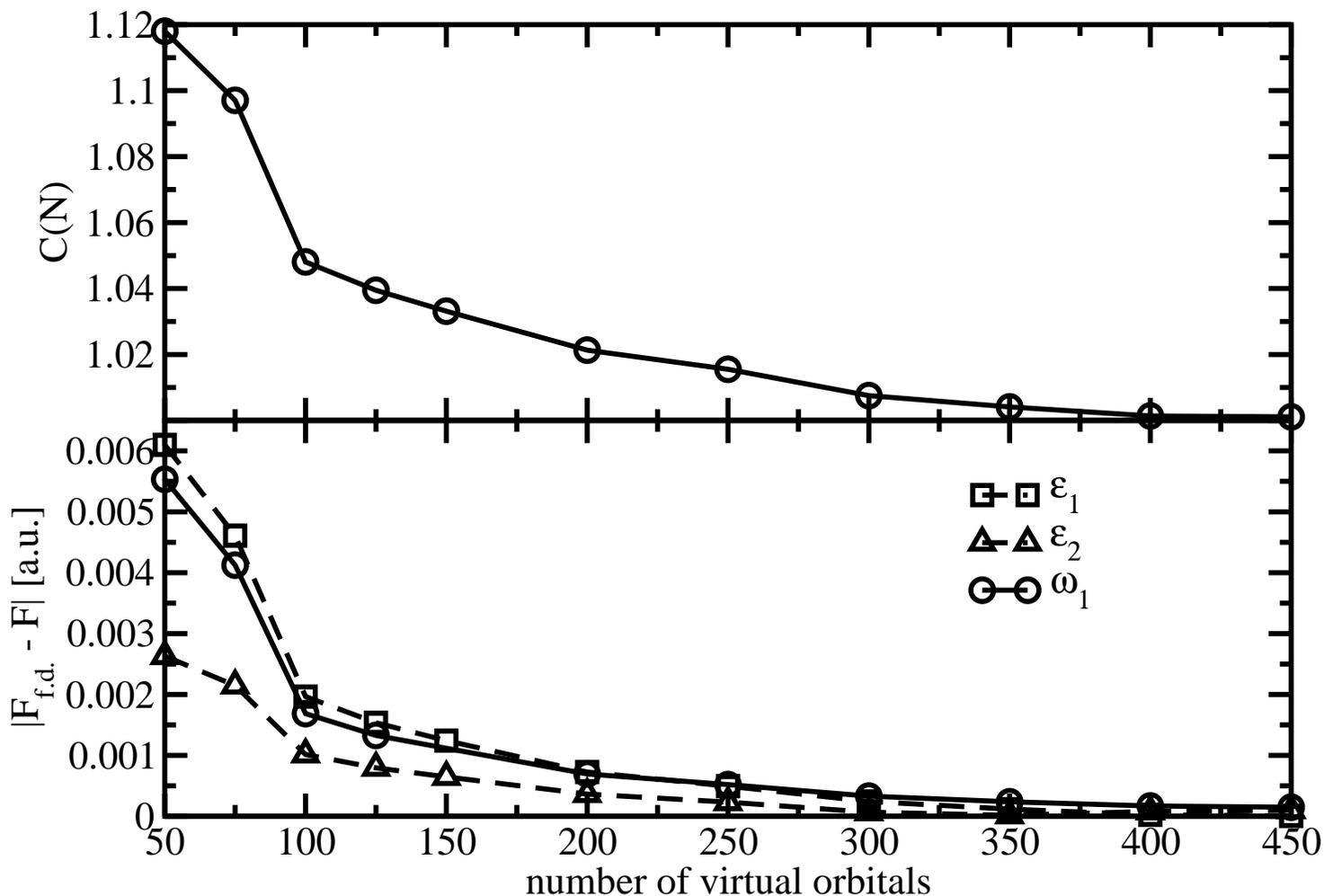}}
\caption{(a) The integral (\ref{complete})
as a function of the number of virtual orbitals included in
the active space.
(b) Absolute deviation of finite difference and analytic derivatives of the 
Kohn-Sham HOMO and LUMO energies of H$_2$ at a bond length of
1.0 a.u. as a function of the number of virtual Kohn-Sham orbitals
included in the active space. 
(c) Absolute deviation of finite difference and analytic derivatives of 
the first singlet excitation energy of  H$_2$ at a bond length of
1.0 a.u. as a function of the number of virtual Kohn-Sham orbitals
included in the active space.
The calculations were carried out
in a cubix box of length 6 a.u. with periodic boundary conditions
and a plane wave cutoff of 40 Ry.}
\label{h2-grad}
\end{figure}

\clearpage
\begin{figure}
\centerline{
\epsfig{figure=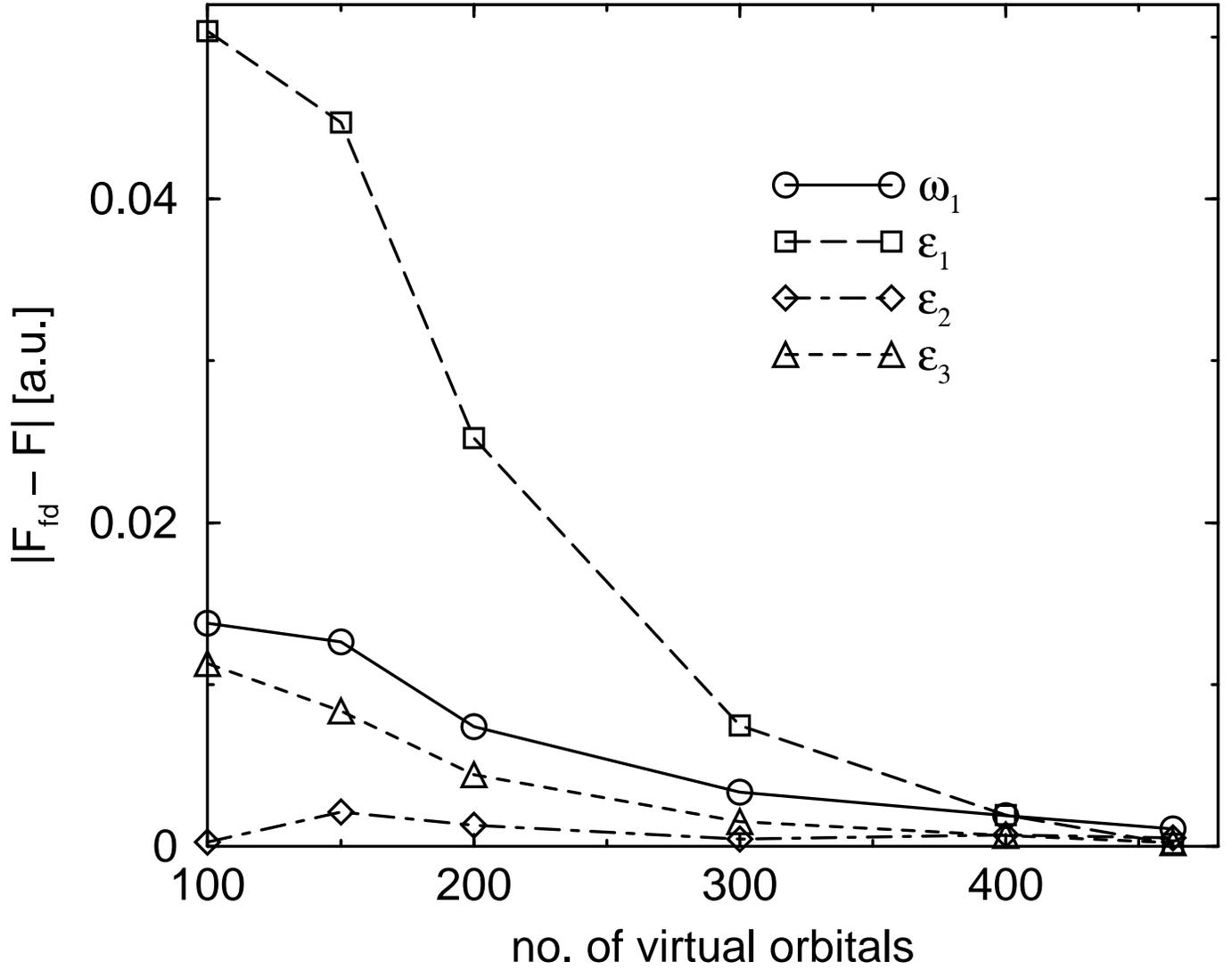,height=0.9\columnwidth}
}
\caption{
Absolute deviation of the analytic derivatives from the
respective finite difference values for the
the first singlet excitation energy, $\omega_1$, as well as
the three lowest KS orbital energies, $\epsilon_i$ ($i=1,2,3$), of  N$_2$ at a bond length of
2.0 a.u. as a function of the number of virtual Kohn-Sham orbitals
included in the active space. The calculations were carried out
in a cubix box of length 6 a.u. with periodic boundary conditions
and a plane wave cutoff of 40 Ry.
}
\label{n2-grad}
\end{figure}

\clearpage
\begin{figure}
\centerline{
\epsfig{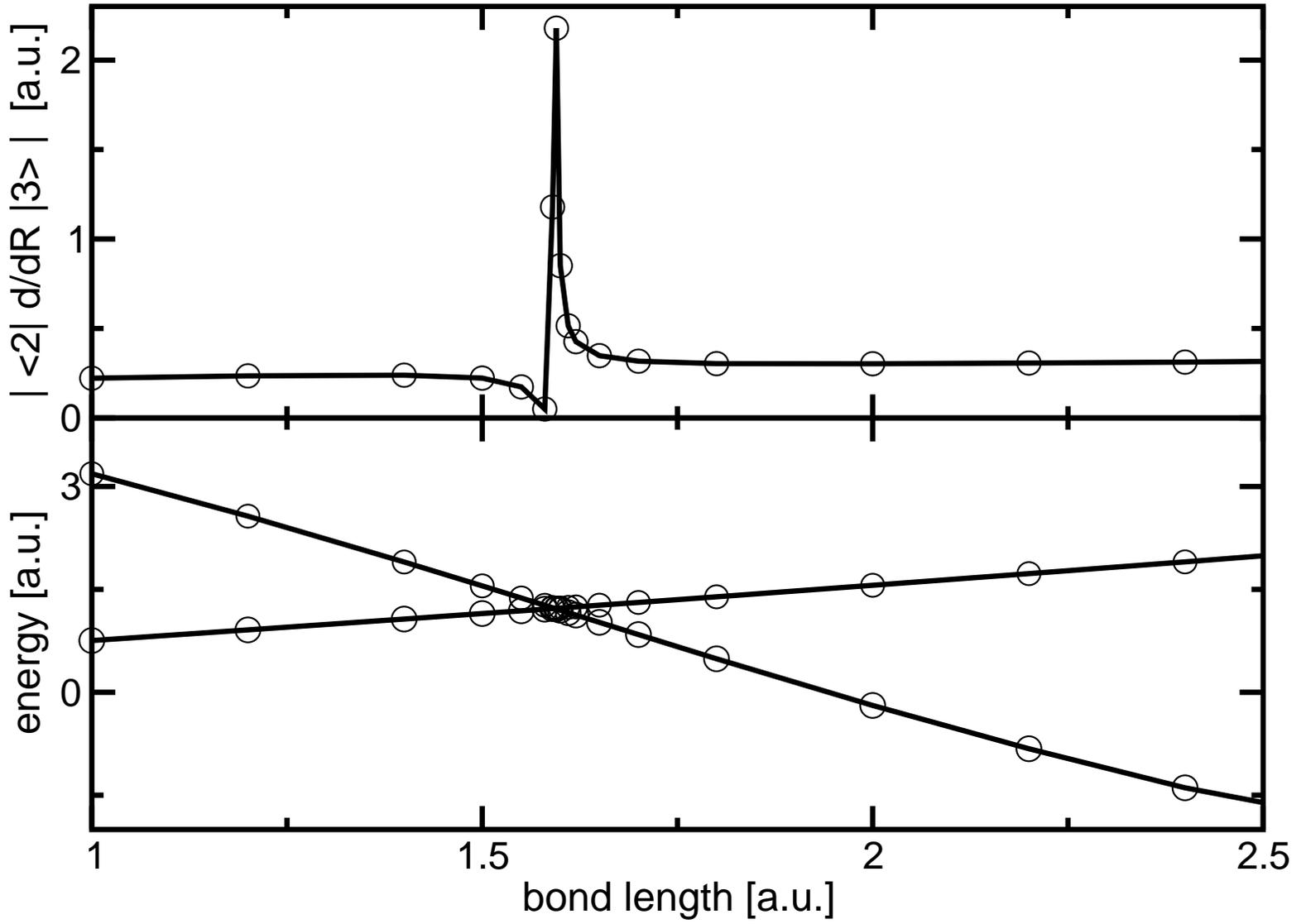}
}
\caption{Upper panel: Absolute value of the nonadiabatic coupling
matrix element between the KS orbitals $\psi_2$ and
$\psi_3$ along the molecular axis of H$_2$ as a function of 
the bond length. Lower panel: KS orbital energies,
$\epsilon_2$ and $\epsilon_3$ of H$_2$  as a function of 
bond length. The calculations were carried out
in a periodic orthorhombic box of size $8\times 5.6 \times 5.6$
a.u.$^3$ using a plane wave cutoff of 40~Ry.}
\end{figure}

\end{document}